\documentclass{article}

     \usepackage[final, nonatbib]{neurips_2021}

\usepackage[utf8]{inputenc} 
\usepackage[T1]{fontenc}    
\usepackage{hyperref}       
\usepackage{url}            
\usepackage{booktabs}       

\usepackage{nicefrac}       
\usepackage{microtype}      

\usepackage{cite}
\usepackage{amsmath,amssymb,amsfonts}
\usepackage{algorithmic}
\usepackage{graphicx}
\usepackage{textcomp}
\usepackage{enumitem}
\usepackage{multirow}
\usepackage{caption}
\usepackage{subcaption}
\usepackage{float}
\usepackage[table,xdraw]{xcolor}
\def\BibTeX{{\rm B\kern-.05em{\sc i\kern-.025em b}\kern-.08em
    T\kern-.1667em\lower.7ex\hbox{E}\kern-.125emX}}
    
\newcommand{\ie}{\textit{i}.\textit{e}., }
\newcommand{\eg}{\textit{e}.\textit{g}. }

\begin{document}

\title{Damage Estimation and Localization from Sparse Aerial Imagery
\thanks{DISTRIBUTION STATEMENT A. Approved for public release. Distribution is unlimited.

This material is based upon work supported by the United States Air Force under Air Force Contract No. FA8702-15-D-0001. Any opinions, findings, conclusions or recommendations expressed in this material are those of the author(s) and do not necessarily reflect the views of the United States Air Force.

© 2021 Massachusetts Institute of Technology.

Delivered to the U.S. Government with Unlimited Rights, as defined in DFARS Part 252.227-7013 or 7014 (Feb 2014). Notwithstanding any copyright notice, U.S. Government rights in this work are defined by DFARS 252.227-7013 or DFARS 252.227-7014 as detailed above. Use of this work other than as specifically authorized by the U.S. Government may violate any copyrights that exist in this work.

This work was also made possible in part due to funding from NSF D-ISN project award \# 2039771.}
}

\author{{Ren\'e Garc\'ia Franceschini}\\
\textit{Institute for Data, Systems and Society} \\
\textit{Massachusetts Institute of Technology}\\
Cambridge, MA \\
ragarcia@mit.edu
\And
{Jeffrey Liu}\\
\textit{MIT Lincoln Laboratory}\\
Lexington, MA \\
Jeffrey.Liu@ll.mit.edu
\And
{Saurabh Amin}\\
\textit{Dept. of Civil and Environmental Engineering} \\
\textit{Massachusetts Institute of Technology}\\
Cambridge, MA \\
amins@mit.edu}

\maketitle

\begin{abstract}
Aerial images provide important situational awareness for responding to natural disasters such as hurricanes. They are well-suited for providing information for damage estimation and localization (DEL); \ie characterizing the type and spatial extent of damage following a disaster. Despite recent advances in sensing and unmanned aerial systems technology, much of post-disaster aerial imagery is still taken by handheld DSLR cameras from small, manned, fixed-wing aircraft.
However, these handheld cameras lack IMU information, and images are taken opportunistically post-event by operators. As such, DEL from such imagery is still a highly manual and time-consuming process. We propose an approach to both detect damage in aerial images and localize it in world coordinates. The approach is based on using structure from motion to relate image coordinates to world coordinates via a projective transformation, using class activation mapping to detect the extent of damage in an image, and applying the projective transformation to localize damage in world coordinates. We evaluate the performance of our approach on post-event data from the 2016 Louisiana floods, and find that our approach achieves a precision of 88\%. Given this high precision using limited data, we argue that this approach is currently viable for fast and effective DEL from handheld aerial imagery for disaster response.

\end{abstract}

\section{Application Context}

Natural disasters, such as hurricanes and floods, can cause major loss of life and property; the intensity, scope, and the frequency of such disasters may be further exacerbated by global climate change \cite{aalst_impacts_2006}. Timely information about the distribution and nature of damage following a disaster can help provide important context and information for emergency managers' decision-making \cite{cova_1999}. Increasingly, satellite and aerial imagery are being incorporated into post-disaster needs assessment~\cite{Maxar_Technologies_2016}. However, while techniques exist for extracting information from orthorectified satellite and aerial imagery~\cite{gupta2019xbd, chen2018benchmark, rahnemoonfar2020floodnet, gupta2020rescuenet,cooner2016detection}, methods for more general aerial imagery (such as oblique imagery from handheld cameras) have received far less attention. This is a critical limitation because post-disaster aerial imagery taken from handheld DSLR cameras from small, manned, fixed-wing aircraft remains popular due to the relatively low cost, high availability, conformity with existing regulations, and existing training programs associated with the practice \cite{department_of_homeland_security_dhsfemapia-055_2020, civil_air_patrol_capabilities_2020}. 
These handheld cameras lack IMU information, and images are taken opportunistically post-event by human operators, resulting in sparsely-sampled images taken at oblique angles. In this paper, we pose the question: \textbf{how can we use aerial imagery from an arbitrary camera setup in order to rapidly and effectively aid in post-disaster situational awareness?}


We focus on a specific component of post-disaster needs assessment, which we refer to as Damage Estimation and Localization (DEL). We broadly define \textit{damage} as an identifiable destruction of an infrastructure component or utility resulting from a specific event (in our case, a natural disaster). We then define \textit{estimation} as the detection of an instance of damage in an image. Finally, \textit{localization} is the act of assigning world coordinates to the estimated instance of damage. Our main contribution is a practically implementable approach that uses sparse, oblique aerial disaster imagery from handheld cameras to carry out DEL. To our knowledge, our approach  is the only one that does estimation without relying on training data that includes bounding boxes or segmented images; and localization without inertial measurement unit (IMU) information or a known geotransform. We show that this method applied to flooding DEL achieves a precision of 88\% when compared against official estimates from the 2016 Louisiana floods. We believe our approach provides emergency managers the means to do fast and effective DEL using existing equipment. 

The approach consists of two stages: a pre-disaster and a post-disaster stage. In the pre-disaster stage, a neural network is trained to classify damage within low altitude disaster imagery. The post-disaster stage is comprised of two parallel 'pipelines', whose outputs are combined at the end. The first pipeline takes a collection of images from an area of interest and reconstructs the scene using structure from motion to obtain a projective transformation that relates image coordinates to world coordinates. The second pipeline takes individual images from the area of interest and produces polygons that cover the extent of the damage that is detected using class activation mapping. The projective transformation is then applied to the damage polygons to produce an estimate of damage locations in world coordinates. Appendix~\ref{app:flowchart} illustrates this approach. 

Existing work in this area has focused primarily on nadir (top-down) or orthorectified imagery. For georeferencing, common approaches use visual feature-based methods such as SIFT to register satellite or nadir drone imagery to other known georeferenced images~\cite{lowe1999object, oh2011automatic, zhuo2017automatic, Goncalves_Corte-Real_Goncalves_2011}. However, methods such as SIFT have been shown to perform poorly under extreme changes in perspective and sensor specifications~\cite{zhuo2017automatic, shetty2019uav}. Deep learning approaches using Siamese neural networks have been proposed for georeferencing aerial images~\cite{tian2017cross, kim2017satellite, shetty2019uav, liu2019lending}; however, these approaches do not recover the orientation information necessary to project the oblique images to the world coordinates. Indeed, most available literature on registering oblique imagery relies on some variant of structure from motion or multiview stereo~\cite{verykokou2018oblique}, which is the approach we take for georeferencing. For detecting damage, various deep learning approaches have achieved high accuracy in object detection challenges such as xView2~\cite{gupta2019xbd, chen2018benchmark, rahnemoonfar2020floodnet}. Previous work has also attempted to overcome lack of training data in either satellite or aerial images via transfer learning from satellite to aerial or vice versa~\cite{cao2016vehicle, liang2016transfer, ji2019scale}. Once again, these approaches were only applied to orthorectified imagery. Given the lack of tools and datasets developed for detecting damage within oblique aerial photos, we pursue a weakly supervised approach to detecting damage using class activation mapping.

\section{Methods} \label{sec:methods}


\subsection{Low Altitude Disaster Imagery dataset} \label{subsec:ladi}
The data used in this paper comes from the Low Altitude Disaster Imagery (LADI) dataset~\cite{liuLargeScale2019}. LADI is a publicly available dataset consisting of images taken by the United States Civil Air Patrol (CAP) in the aftermath of natural disasters, and annotated by crowdsourced workers with hierarchical image-level labels representing five broad categories: Damage, Environment, Infrastructure, Vehicles, and Water. Within each category, there are a number of more specific annotation labels. We focus on the \textit{flooding/water damage} label within in the ``Damage" category. 


\subsection{Damage estimation within an image}\label{subsec:damage_detection}

This section describes our damage estimation pipeline. We train an image classifier to detect damage, and use the class activation map from the classifier to estimate the location of damage within images.

\subsubsection{Classification with ResNet}
\label{subsubsec:damage_categories}

To perform the image labeling task, we use a ResNet-50 backbone architecture~\cite{he2016deep} pretrained on ImageNet~\cite{deng2009imagenet} and fine tuned on the LADI dataset. ImageNet weights are used due to their status the default pretrained weights for popular implementations of ResNet-50~\cite{team_keras_nodate,noauthor_torchvisionmodels_nodate}; additional work is necessary to determine if performance could be improved by pretraining on an overhead imagery dataset, such as xBD~\cite{gupta2019xbd}.  Images within the LADI dataset were shown to multiple workers per category; we considered an image a positive example if at least two workers marked it as such. Appendix \ref{app:labeling} provides further details on how we constructed the training labels from worker responses.
We split the dataset 80\%/10\%/10\% for the training, validation and testing sets, respectively. The ResNet achieves a performance of 77\% accuracy, 65\% precision, and 79\% recall on the validation set; see Appendix \ref{app:labeling} for more details.



\begin{figure}
\centering
\begin{subfigure}{.27\linewidth}
  \centering
  \includegraphics[width=.99\linewidth]{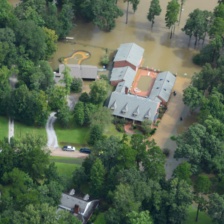}  
  \caption{Original image}
  \label{fig:poly_1}
\end{subfigure}
\begin{subfigure}{.27\linewidth}
  \centering
  \includegraphics[width=.99\linewidth]{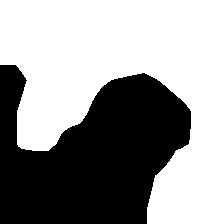}  
  \caption{CAM mask}
  \label{fig:poly_2}
\end{subfigure}
\begin{subfigure}{.27\linewidth}
  \centering
  \includegraphics[width=.99\linewidth]{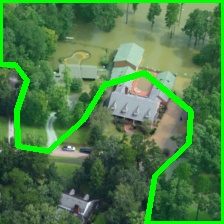}  
  \caption{Polygon tracing}
  \label{fig:poly_3}
\end{subfigure}

\caption{Stages of our polygon tracing approach.}
\label{fig:poly_tracing}
\end{figure}

\subsubsection{Class activation mapping and polygon tracing}
After training, we utilize the class activation mapping (CAM) approach from Zhou et al.~\cite{zhou2016learning} to localize the extent of the detected class within the image. CAM is a technique for weakly-supervised object detection, which leverages the average pooling layer at the end of the ResNet architecture to detect areas within an image that are important for classifying a particular class. We threshold the pooling layer (see Appendix~\ref{app:cam_details} to generate a mask, and convert the mask into a set of polygons using~\cite{suzuki1985topological}. Fig.~\ref{fig:poly_tracing} shows different stages of this procedure.


\subsection{Damage localization}
This section describes the localization pipeline: estimating the projective transformation relating image and world coordinates using SfM, and applying it to transform the flooding polygons.

\label{subsec:damage_localization}
\subsubsection{Reconstruction using structure from motion}
Structure from motion is a technique that, using images from a camera moving through an environment, can produce a point cloud of the environment~\cite{andrew2001multiple}. By taking advantage of the GPS tags from the image metadata or from outside sensors, structure from motion has been used to create inexpensive, georeferenced elevation models from drone and aircraft imagery~\cite{fonstad2013topographic}. We use this technique as an intermediate step to obtaining the projective transformation that relates image coordinates to world coordinates. We base our implementation off the well-known OpenSfM library, an open source library for structure from motion~\cite{mapillary/OpenSfM_2021}.

Since fixed wing aircraft tend to fly in a straight line, and have a relatively large turn radius compared to rotary-wing aircraft, sequential images collected from fixed-wing platforms tend to be approximately collinear. Thus, some reconstructions have an additional degree of freedom from rotating about the line that goes through the GPS coordinates. We address this by separately estimating the direction of the up-vector (the vector opposite the direction of gravity) and enforcing it in the reconstruction. Previous implementations in urban environments have suggested estimating vanishing points to estimate the up-vector~\cite{wang2013accurate}. This may be difficult if there are few straight features (such as roads), which is common in rural areas.

To estimate the up-vector, we assume the ground is approximately flat. We first fit a plane through the reconstructed features using RANSAC~\cite{fischler1981random}. There is a pair of possible antiparallel unit normal vectors to this plane, one of which is the up-vector. Because of the aerial nature of the data, the location of the images must be above the ground plane. Therefore, we choose the vector that has a positive projection onto the image location in East, North Up (ENU) coordinates and denote it $v_{up}$. Finally, we rotate the reconstruction so that $v_{up}$ indeed points upwards. Specifically, we rotate it by $R_z$ such that $R_z v_{up} = \hat{z}$ when it is initialized, and the up-vector is enforced during bundle adjustment. In areas with significant variation in local topology, a digital elevation model (DEM) can be used to estimate the ground plane. We had initially incorporated a digital elevation model (DEM). However, since the region that we considered is relatively flat, there was no difference in performance, and thus, we do not report the DEM results.

\subsubsection{Image-to-world projective transformation} \label{subsubsec:proj_trans}
The final step in our georeferencing pipeline is estimating the transformation from image coordinates to world coordinates, and applying it to the detected damage polygons. As discussed previously, the images are of mostly flat surfaces, and thus the coordinates can be related by a projective transformation~\cite{andrew2001multiple}, and outliers can be filtered through RANSAC~\cite{fischler1981random}. Of all of the images that were reconstructed using OpenSfM, we retained those where at least 20\% of matches between image and world coordinates were inliers. Of the retained images, we found that some images produced extremely large image footprints (\ie, the projection of the image edges onto the ground). We identified that these photos were those that were so oblique that the horizon was visible. Because such images require more complex transformations, we eliminated them from our analysis. We considered two heuristic criteria for eliminating such images. First, we eliminated images whose total area were greater than some value $\gamma_1$. Second, we did not consider images where the ratio of the longest side to the shortest side of the minimum area rectangle that covered the entire footprint was greater than $\gamma_2$. We report our results for a variety of $\gamma_1$ and $\gamma_2$ parameters to illustrate the effectiveness of our approach.

\section{Evaluation and Results} \label{sec:results}
\begin{figure}
\begin{center}
   \includegraphics[width=0.35\linewidth]{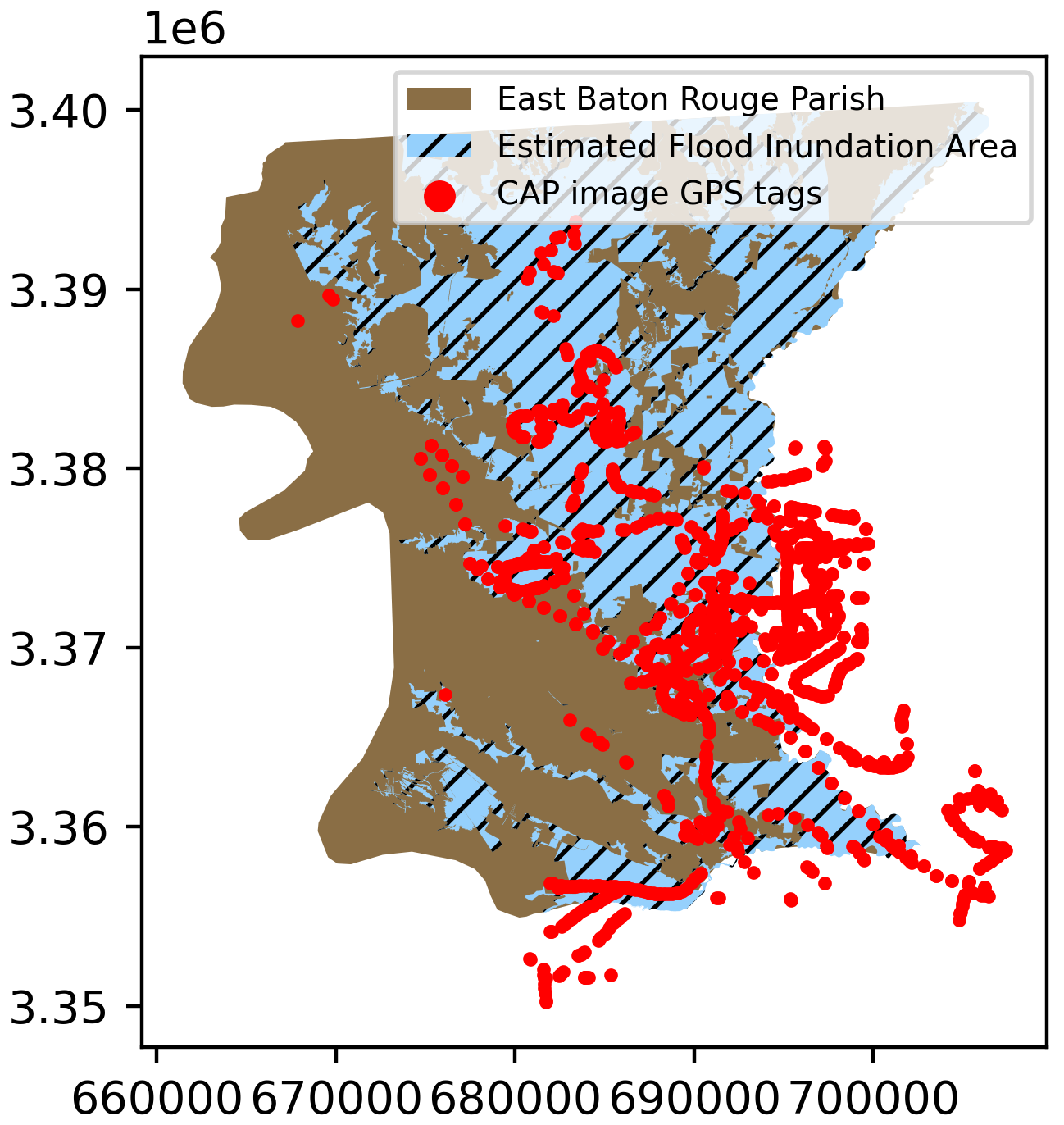}
\end{center}
   \caption{Map of East Baton Rouge parish and image GPS tags.}
\label{fig:flooding}
\end{figure}

In this section, we evaluate the performance of our approach at DEL using images from the 2016 Louisiana floods. Figure~\ref{fig:flooding} shows the administrative boundary of the East Baton Rouge parish in Louisiana, the parish's estimated flood inundation area~\cite{city_2016}, and the coordinates of all CAP image with GPS locations within 5 km of the administrative boundary. In total, the flooding event covered 536 km\textsuperscript{2} (44\% of the total area of the parish). Our analysis includes 1615 CAP images that were taken in August 2016 immediately after the flooding event. Of these CAP images that were considered, 809 were successfully reconstructed by OpenSfM. At the same time, 996 of the images were identified as having flooding. Finally, 559 images both completed the georeferencing pipeline \textit{and} were identified as flooding. Additional images were then filtered based on the criteria described in Section~\ref{subsubsec:proj_trans}.

We evaluated three different methods of estimating flooding. First, we use the GPS tag of these images as a baseline, where we calculate the precision as the proportion of the flood images that lie in the FEMA estimates. Second, we estimate the flooding using the entire footprints of images classified as containing \textit{``flooding/water"} (SfM + binary classification). Finally, we consider our approach of both (SfM + CAM) as the flood estimate. For the SfM + binary classification and SfM + CAM approaches, we compute precision as the fraction of the computed flooding polygons which overlap with the official FEMA estimates. We only consider the flooding within the East Baton Rouge administrative boundary, since we do not have data on flood extent outside of the boundary.

\begin{figure*}[h]
\centering
\begin{subfigure}{.32\linewidth}
  \centering
  \includegraphics[width=.99\linewidth]{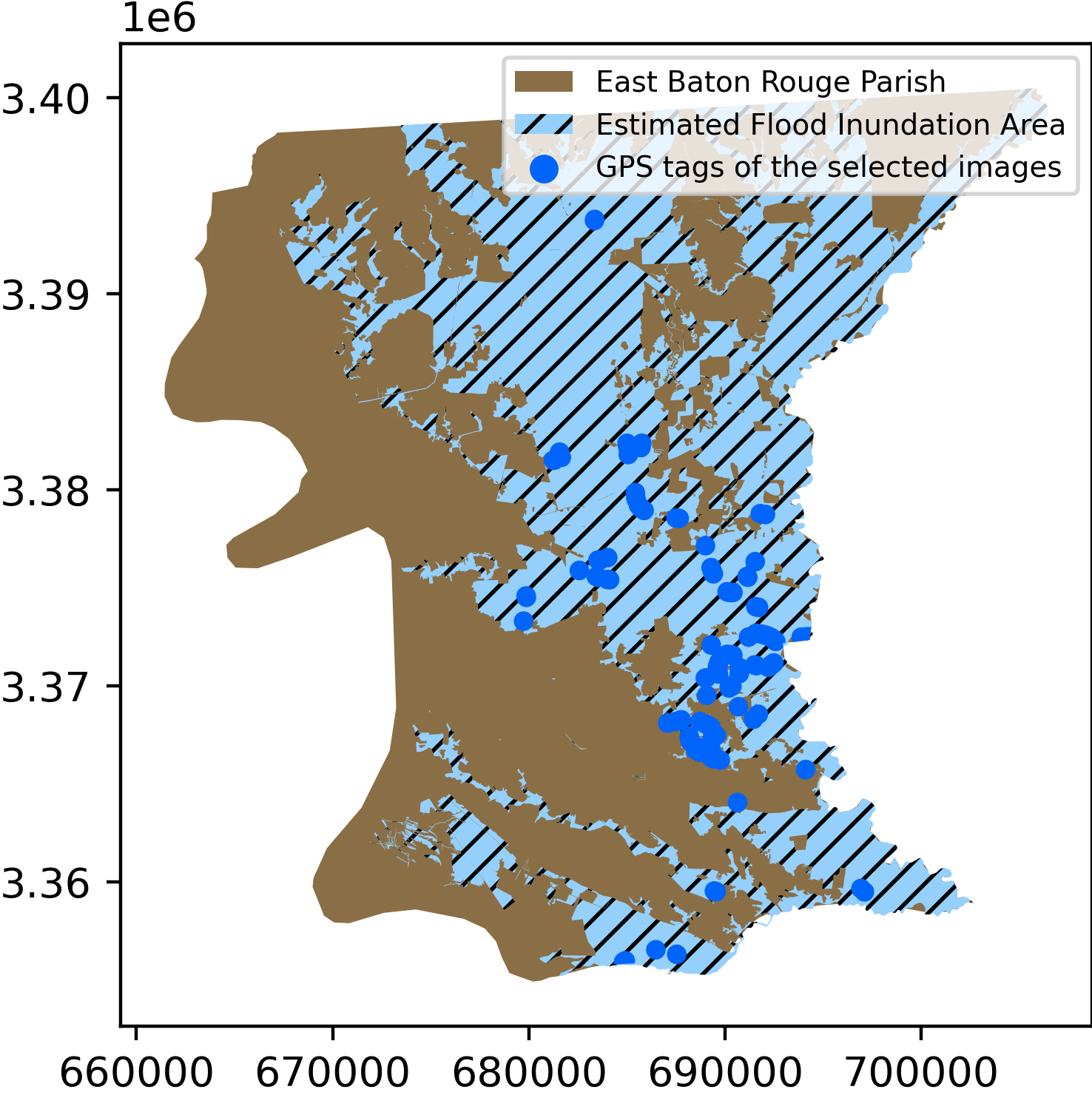}  
  \caption{GPS tags (Prec: 52\%)}
  \label{fig:results_1}
\end{subfigure}
\begin{subfigure}{.32\linewidth}
  \centering
  \includegraphics[width=.99\linewidth]{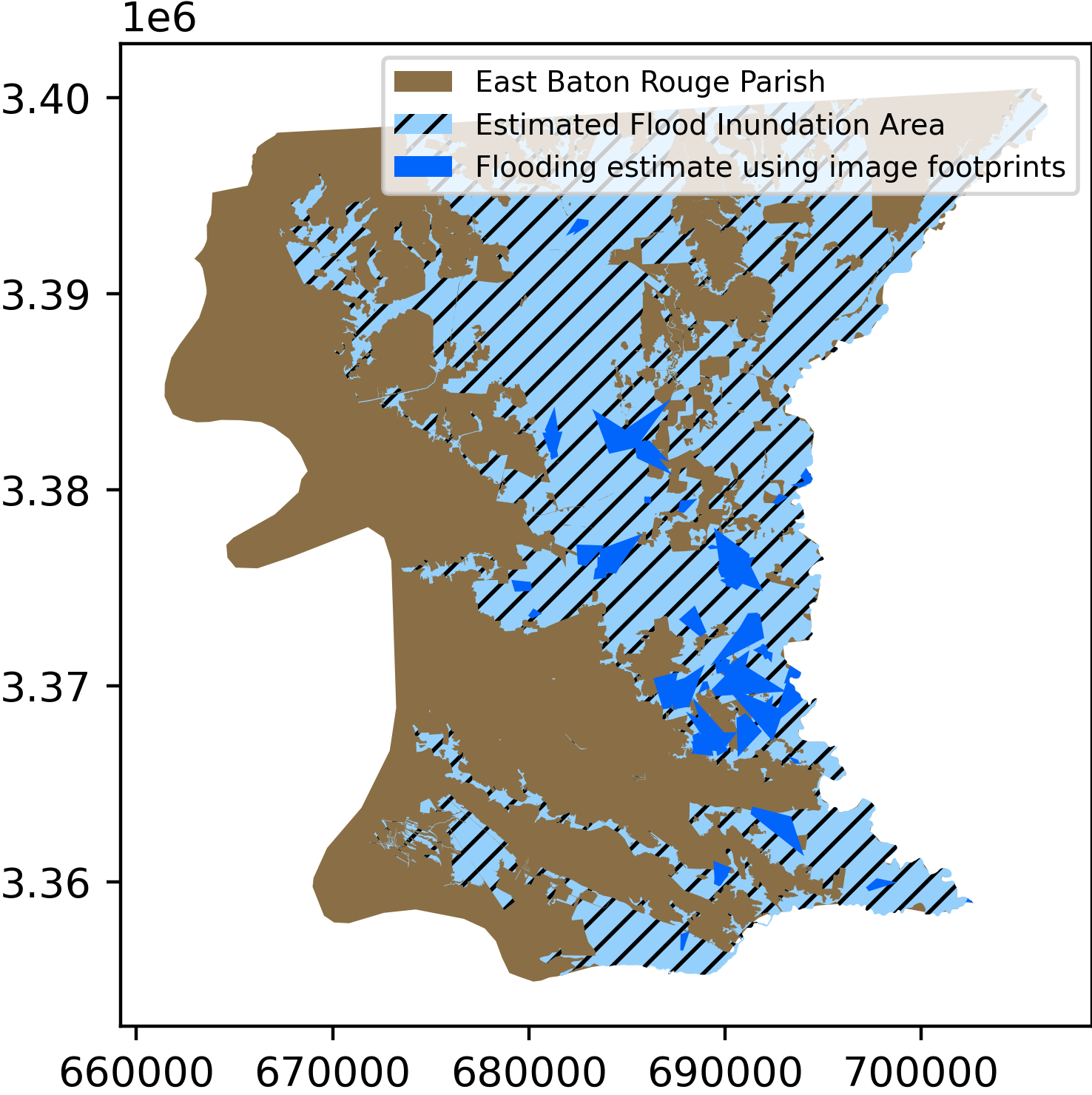}  
  \caption{Image footprints (Prec: 80\%)}
  \label{fig:results_2}
\end{subfigure}
\begin{subfigure}{.32\linewidth}
  \centering
  \includegraphics[width=.99\linewidth]{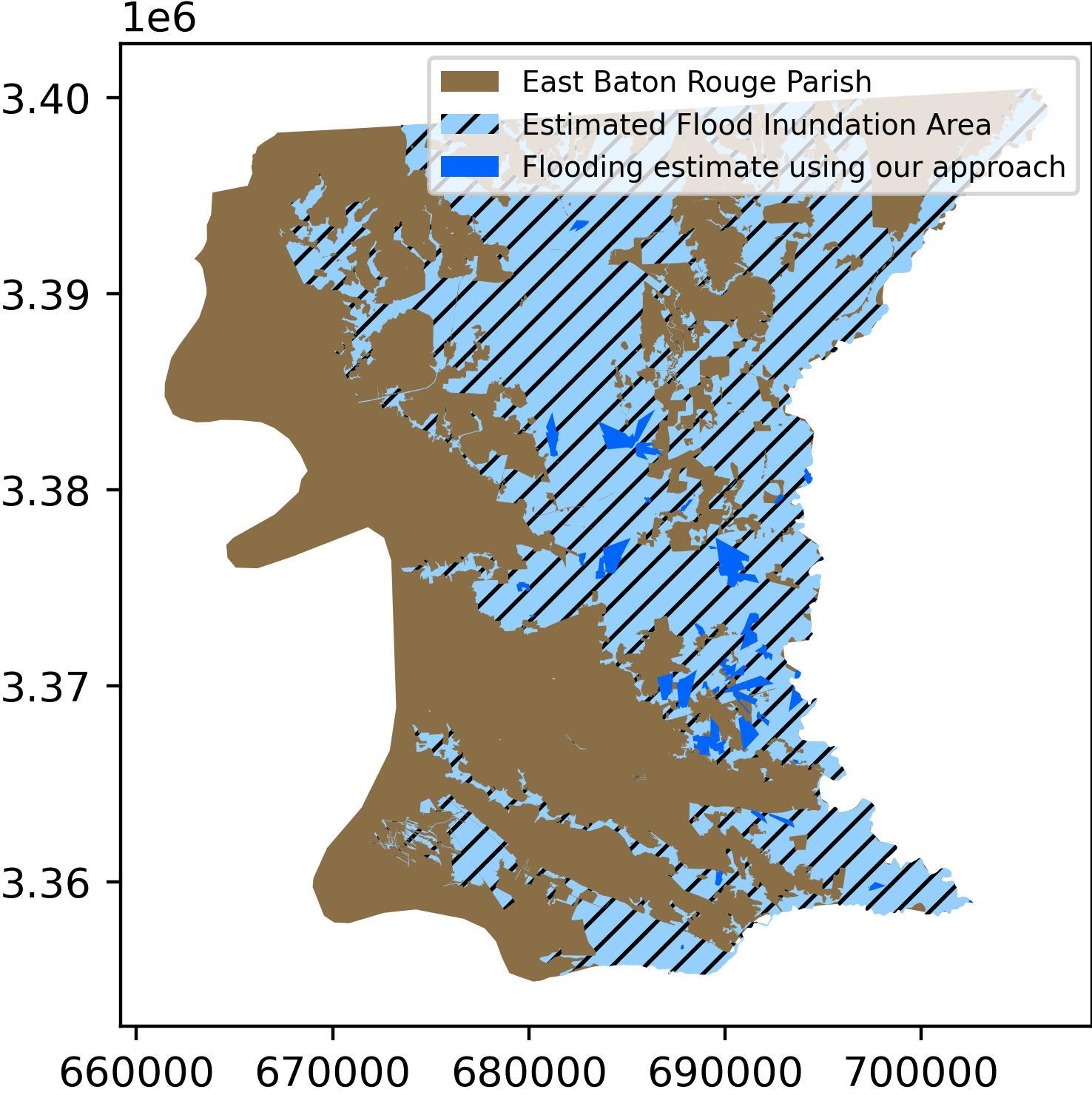}  
  \caption{Our approach (Prec: 88\%)}
  \label{fig:results_3}
\end{subfigure}

\caption{Flooding estimates and precision values using: (a) GPS tags, (b) image footprints (SfM + binary classification), and (c) our approach (SfM + CAM).}
\label{fig:flooding_results}
\end{figure*}


Figure~\ref{fig:flooding_results} shows the different flooding estimates overlaid against the official estimates, as well as the precision values of each method. Note that the precision we report is compared against the flooding extent estimates, and not with ground truth mask of the images (which are not available in the dataset). The reported precision includes both the contributions of CAM and SfM. These estimates were made with $\gamma_1 = 4$ and $\gamma_2= 5$ km\textsuperscript{2}, so that ultimately 243 images were used. These results show a clear improvement going from using the image GPS locations to using the georeferenced footprint. This suggests that using the GPS of the image location is insufficient, as a large number of images containing flooding were taken over areas that were not flooded, and vice versa. Furthermore, we see that our approach (SfM + CAM) further improves the precision compared to using the full image footprint. Especially at the edges of the flooding extent, our method provides a more precise outline of the official estimates than using the full image footprints. Appendix~\ref{app:closeup} provides a closeup view.

\begin{table}[h]
\begin{minipage}{.5\linewidth}
\centering
\caption*{\textbf{SfM + binary classification}}
\begin{tabular}{cccccc}
\multicolumn{1}{l}{}                                       & \multicolumn{1}{l}{}                                    & \multicolumn{4}{c}{$\gamma_2$ (km\textsuperscript{2})}                                                                                                                                                                                     \\ \cline{3-6} 
\multicolumn{1}{l}{}                                       & \multicolumn{1}{l|}{}                                   & \multicolumn{1}{c|}{\cellcolor[HTML]{D9D9D9}\textbf{1}} & \multicolumn{1}{c|}{\cellcolor[HTML]{D9D9D9}\textbf{2.5}} & \multicolumn{1}{c|}{\cellcolor[HTML]{D9D9D9}\textbf{5}} & \multicolumn{1}{c|}{\cellcolor[HTML]{D9D9D9}\textbf{10}} \\ \cline{2-6} 
\multicolumn{1}{c|}{}                                      & \multicolumn{1}{c|}{\cellcolor[HTML]{D9D9D9}\textbf{2}} & \multicolumn{1}{c|}{86}                                  & \multicolumn{1}{c|}{85}                                    & \multicolumn{1}{c|}{85}                                  & \multicolumn{1}{c|}{87}                                   \\ \cline{2-6} 
\multicolumn{1}{c|}{}                                      & \multicolumn{1}{c|}{\cellcolor[HTML]{D9D9D9}\textbf{3}} & \multicolumn{1}{c|}{85}                                  & \multicolumn{1}{c|}{84}                                    & \multicolumn{1}{c|}{85}                                  & \multicolumn{1}{c|}{86}                                   \\ \cline{2-6} 
\multicolumn{1}{c|}{}                                      & \multicolumn{1}{c|}{\cellcolor[HTML]{D9D9D9}\textbf{4}} & \multicolumn{1}{c|}{85}                                  & \multicolumn{1}{c|}{84}                                    & \multicolumn{1}{c|}{80}                                  & \multicolumn{1}{c|}{72}                                  \\ \cline{2-6} 
\multicolumn{1}{c|}{\multirow{-4}{*}{\begin{tabular}[c]{@{}c@{}}$\gamma_1$\\ (unitless)\end{tabular}}}  & \multicolumn{1}{c|}{\cellcolor[HTML]{D9D9D9}\textbf{5}} & \multicolumn{1}{c|}{85}                                  & \multicolumn{1}{c|}{84}                                    & \multicolumn{1}{c|}{77}                                  & \multicolumn{1}{c|}{68}                                  \\ \cline{2-6} 
\end{tabular}
\end{minipage}
\begin{minipage}{.5\linewidth}
\centering
\caption*{\textbf{SfM + CAM}}
\begin{tabular}{cccccc}
\multicolumn{1}{l}{}                                       & \multicolumn{1}{l}{}                                    & \multicolumn{4}{c}{$\gamma_2$ (km\textsuperscript{2})}                                                                                                                                                                                     \\ \cline{3-6} 
\multicolumn{1}{l}{}                                       & \multicolumn{1}{l|}{}                                   & \multicolumn{1}{c|}{\cellcolor[HTML]{D9D9D9}\textbf{1}} & \multicolumn{1}{c|}{\cellcolor[HTML]{D9D9D9}\textbf{2.5}} & \multicolumn{1}{c|}{\cellcolor[HTML]{D9D9D9}\textbf{5}} & \multicolumn{1}{c|}{\cellcolor[HTML]{D9D9D9}\textbf{10}} \\ \cline{2-6} 
\multicolumn{1}{c|}{}                                      & \multicolumn{1}{c|}{\cellcolor[HTML]{D9D9D9}\textbf{2}} & \multicolumn{1}{c|}{90}                                  & \multicolumn{1}{c|}{89}                                    & \multicolumn{1}{c|}{89}                                  & \multicolumn{1}{c|}{89}                                   \\ \cline{2-6} 
\multicolumn{1}{c|}{}                                      & \multicolumn{1}{c|}{\cellcolor[HTML]{D9D9D9}\textbf{3}} & \multicolumn{1}{c|}{90}                                  & \multicolumn{1}{c|}{88}                                    & \multicolumn{1}{c|}{87}                                  & \multicolumn{1}{c|}{87}                                   \\ \cline{2-6} 
\multicolumn{1}{c|}{}                                      & \multicolumn{1}{c|}{\cellcolor[HTML]{D9D9D9}\textbf{4}} & \multicolumn{1}{c|}{90}                                  & \multicolumn{1}{c|}{88}                                    & \multicolumn{1}{c|}{88}                                  & \multicolumn{1}{c|}{87}                                  \\ \cline{2-6} 
\multicolumn{1}{c|}{\multirow{-4}{*}{\begin{tabular}[c]{@{}c@{}}$\gamma_1$\\ (unitless)\end{tabular}}}  & \multicolumn{1}{c|}{\cellcolor[HTML]{D9D9D9}\textbf{5}} & \multicolumn{1}{c|}{90}                                  & \multicolumn{1}{c|}{88}                                    & \multicolumn{1}{c|}{87}                                  & \multicolumn{1}{c|}{87}                                  \\ \cline{2-6} 
\end{tabular}
\end{minipage}
\vspace{2pt}
\caption{Precision for both approaches in percent for various values of $\gamma_1$ and $\gamma_2$.}
\label{tab:precision_ours}
\end{table}

To characterize the improvement that our approach (SfM+CAM) provides over using the entire image footprints (SfM + binary classification), we compute in Table~\ref{tab:precision_ours} the precision for various combinations of $\gamma_1$ and $\gamma_2$. We see that for all chosen combinations, the SfM+CAM approach is more precise. In addition, the precision of the footprint approach degrades from 86\% to 68\% as the values of $\gamma_1$ and $\gamma_2$ increase, while the performance of our approach only decreases from 90\% to 87\%. Thus, our approach is more robust to the choices of these parameters, and allows us to incorporate more of the highly-oblique images that would otherwise affect performance in the footprint approach. 


\section{Discussion and Conclusion} \label{sec:conclusion}

In this paper, we addressed the problem of Damage Estimation and Localization (DEL) using aerial images taken from handheld cameras in small, manned, fixed-wing aircraft. While this mode of collecting imagery is cost-efficient and compliant with existing regulations, such images are often highly oblique, sparsely and irregularly sampled, and lack IMU information. We proposed an approach to performing DEL from such images by combining SfM and CAM to georeference images and detect damage within them. When compared against official flooding estimates from the 2016 Louisiana floods, our approach achieved a precision of 88\%, which outperforms the naive approaches of using the image GPS locations or image footprint. Our approach faces some limitations: extremely oblique images which contain the horizon are unable to be georeferenced using projective transformations, and images that do not overlap with other images cannot be processed using SfM. Nevertheless, our approach can quickly generate an estimate of damage distribution from aerial imagery that is already being collected with existing equipment.

\bibliographystyle{IEEEtran}
{\small
\bibliography{egbib}}

\newpage
\appendix
\section{Flowchart}
\label{app:flowchart}
\begin{figure}[h!]
\begin{center}
   \includegraphics[width=0.5\linewidth]{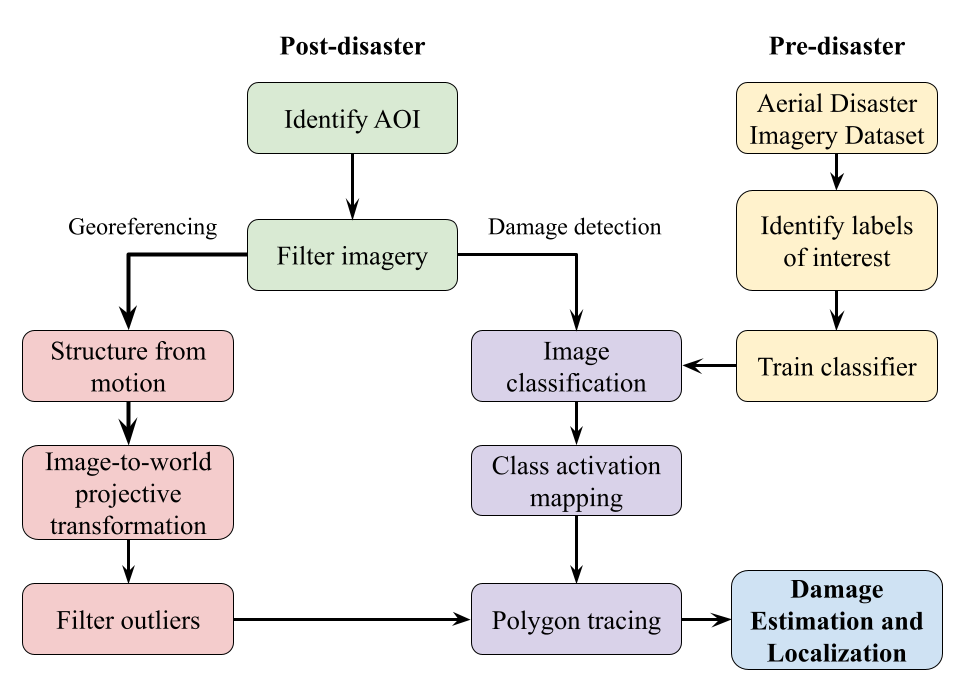}
\end{center}
   \caption{Flowchart depicting our approach.}
\label{fig:flowchart}
\end{figure}

\section{Labeling scheme}
\label{app:labeling}
In the construction of the LADI dataset, images were shown to a variable number of workers (generally between 3-5); each worker was asked to identify which, if any, labels for a given category (e.g. ``Damage") applied to that image \cite{liuLargeScale2019}. Responses from all workers were recorded. As a result, images could have potentially differing annotations between workers due to factors such as subjectivity (worker did not think the label applied) or error (incorrect input). To account for these conflicts, we designate three different labelling schemes for classifier training and evaluation:

\begin{enumerate}[label=\Alph*)]
    \item $B_{i, j} > 1$,
    \item $B_{i, j} > 2$,
    \item $B_{i, j} > 1$ and $B_{i, j}/w_i > \underset{\forall i}{\mathrm{median}} \{B_{i, j}/w_i\}$,
\end{enumerate}

where $B_{i, j}$ is the number of workers that labelled image $i$ as class $j$ and $w_i$ is the number of workers that labelled image $i$ at all. We trained and tested using different combinations of these labelling scheme to determine a balance between filtering out noise and preserving a sufficiently representative dataset.

\begin{table}[h!]
\centering
\begin{tabular}{|c|c|c|c|c|}
\hline
\textbf{\begin{tabular}[c]{@{}c@{}}Train \\ label\end{tabular}} & \textbf{\begin{tabular}[c]{@{}c@{}}Test\\ label\end{tabular}} & \textbf{\begin{tabular}[c]{@{}c@{}}Accuracy\\ (\%)\end{tabular}} & \textbf{\begin{tabular}[c]{@{}c@{}}Precision\\ (\%)\end{tabular}} & \textbf{\begin{tabular}[c]{@{}c@{}}Recall\\ (\%)\end{tabular}} \\ \hline
\multirow{3}{*}{A}                                                     & A                                                                    & 77                                                               & 65                                                                & 79                                                             \\ \cline{2-5} 
                                                                       & B                                                                    & 71                                                               & 38                                                                & 91                                                             \\ \cline{2-5} 
                                                                       & C                                                                    & 70                                                               & 42                                                                & 83                                                             \\ \hline
\multirow{3}{*}{B}                                                     & A                                                                    & 77                                                               & 75                                                                & 53                                                             \\ \cline{2-5} 
                                                                       & B                                                                    & 83                                                               & 54                                                                & 68                                                             \\ \cline{2-5} 
                                                                       & C                                                                    & 80                                                               & 55                                                                & 64                                                             \\ \hline
\multirow{3}{*}{C}                                                     & A                                                                    & 79                                                               & 73                                                                & 65                                                             \\ \cline{2-5} 
                                                                       & B                                                                    & 80                                                               & 48                                                                & 80                                                             \\ \cline{2-5} 
                                                                       & C                                                                    & 83                                                               & 53                                                                & 69                                                             \\ \hline
\end{tabular}
\caption{Accuracy, precision and recall values for the three ResNet models. \textit{Train label} refers to the set of labels used in training, while \textit{Test label} refers to the set of labels against which each model was evaluated.}
\label{tab:classification_results}
\end{table}

Table~\ref{tab:classification_results} shows the testing accuracy, precision and recall values for the three ResNet50 classifiers that were trained (one for each training labelling scheme defined above. In order to properly compare the three models, each of the three classifiers was also evaluated against the remaining two labelling schemes. Regardless of the labelling, the actual images that comprised the testing set (as well as the training and validation sets) were the same for all three schemes. For the purposes of this paper, we will refer to each of the models according to their training labelling scheme.

Unsurprisingly, each of three models had the highest accuracy when compared against the labelling scheme they were trained on. With the other two metrics, though, there are noticeable trends. In terms of precision, model B had the highest precision when evaluated against any labelling scheme, followed by C and finally A. With recall, the opposite holds: A has the highest recall across the board, followed by C and then B. These trends are not difficult to justify, since B necessarily has a higher standard for classification as flooding than A. For flooding, C is a compromise between the most lenient labelling scheme (A) and the strictest one (C). In this particular application, we consider false positives to be less serious than false negatives; that is, we would rather think that someone was in danger from flooding when they are not (false positive) than think that they are not in danger when they are (false negative). As such, we proceed using model A for the remainder of the section.

\section{CAM details}
\label{app:cam_details}
Using the terminology in~\cite{zhou2016learning}, the output on the final fully connected layer is given by:

\begin{equation}
    S_c = \sum_k w_k^c \sum_{x, y} f_k(x, y) = \sum_{x, y} \sum_k  w_k^c f_k(x, y),
\end{equation}

where $w_k^c$ is the weight corresponding to class $c$ for the $k$-th unit within conv5 (the final block within ResNet-50) and $f_k(x, y)$ is the activation of the same $k$-th unit at location ($x$, $y$) (such that $\sum_{x, y} f_k(x, y)$ is the output of the global pooling layer). Let $M_c(x, y) = \sum_k  w_k^c f_k(x, y)$, so that:

\begin{equation}
    S_c =  \sum_{x, y} M_c(x, y).
\end{equation}

Here, $M_c(x, y)$ can be viewed as a measure of importance of a spatial coordinate ($x$, $y$) for the class $c=\text{\textit{flooding/water damage}}$, and hence referred to the class activation map. In order to determine the boundaries of the flooding instances, we threshold on $M_c$:

\begin{equation}
    M^{\text{mask}}_c(x, y) = 
        \begin{cases}
            1 &\mbox{if } M_c(x, y) \geq 0 \\
            0 &\mbox{otherwise}
        \end{cases}
\end{equation}

\section{CAM qualitative evaluation}

\begin{figure}[h!]
\centering
\begin{subfigure}{.28\linewidth}
  \centering
  \includegraphics[width=.99\linewidth]{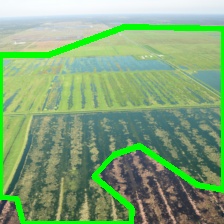}  
  \caption{}
  \label{fig:random_1}
\end{subfigure}
\begin{subfigure}{.28\linewidth}
  \centering
  \includegraphics[width=.99\linewidth]{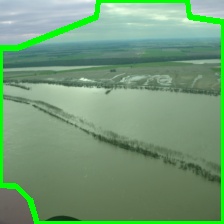}  
  \caption{}
  \label{fig:random_3}
\end{subfigure}
\begin{subfigure}{.28\linewidth}
  \centering
  \includegraphics[width=.99\linewidth]{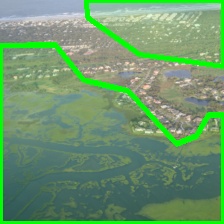}  
  \caption{}
  \label{fig:random_4}
\end{subfigure}

\begin{subfigure}{.28\linewidth}
  \centering
  \includegraphics[width=.99\linewidth]{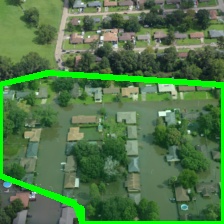}  
  \caption{}
  \label{fig:la_1}
\end{subfigure}
\begin{subfigure}{.28\linewidth}
  \centering
  \includegraphics[width=.99\linewidth]{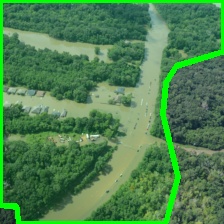}  
  \caption{}
  \label{fig:la_3}
\end{subfigure}
\begin{subfigure}{.28\linewidth}
  \centering
  \includegraphics[width=.99\linewidth]{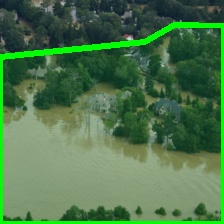}  
  \caption{}
  \label{fig:la_4}
\end{subfigure}

\caption{Sample of CAP images identified as flooding by ResNet model and their associated damage polygons. }
\label{fig:cam_random}
\end{figure}

Figure~\ref{fig:cam_random} shows a sample of LADI images that were classified as having \emph{flooding/water damage}, along with the estimated extent of flooding for qualitative evaluation. We can see that the CAM provides a decent coarse estimate of the extent of water in the image. Flooding is slightly more complicated. While Figures~\ref{fig:random_1} and \ref{fig:random_3} very clearly show flooding events, the top portion of Figure~\ref{fig:random_4} seems to simply be picking up the shoreline. As an important note, we noticed that many images that show large bodies of water also tend to include the horizon in the flooding polygon (\eg Figure~\ref{fig:random_3}). This might be because flooding typically covers a large portion of area, and therefore images that include the horizon might be more likely to also include flooding. This underscores the importance of filtering images with large footprints after georeferencing. 
Figures~\ref{fig:la_1}, \ref{fig:la_3} and \ref{fig:la_4} are images that were identified as flooding from the Louisiana 2016 floods. Even in this case where many images have large portions of flooding, our approach is still able to trace the extent of the water.

\section{Closeup comparison}
\begin{figure}[h!]
     \centering
     \begin{subfigure}[b]{0.45\linewidth}
         \centering
         \includegraphics[width=\linewidth]{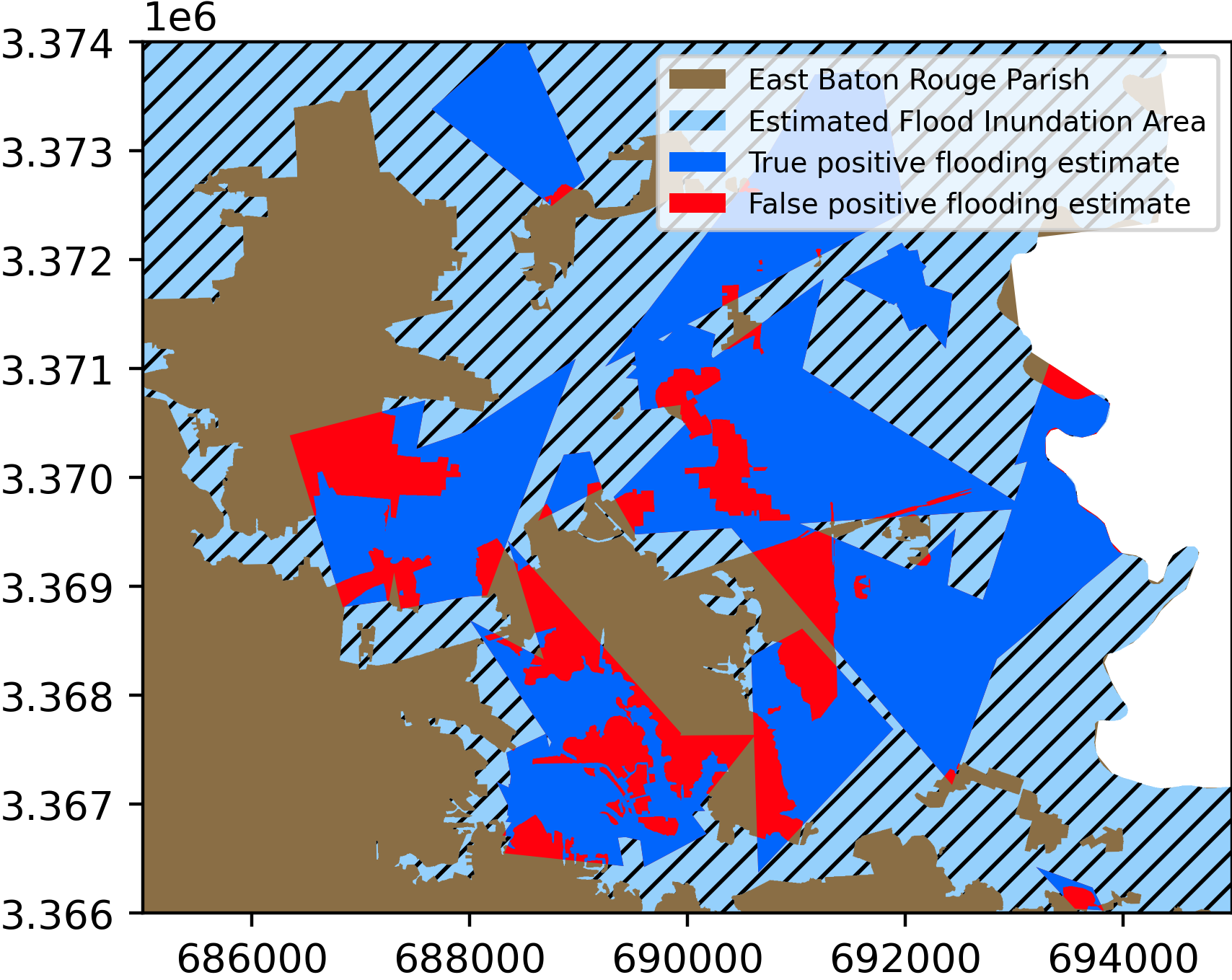}
         \caption{Image footprint}
         \label{fig:footprint_closeup}
     \end{subfigure}
     \begin{subfigure}[b]{0.45\linewidth}
         \centering
         \includegraphics[width=\linewidth]{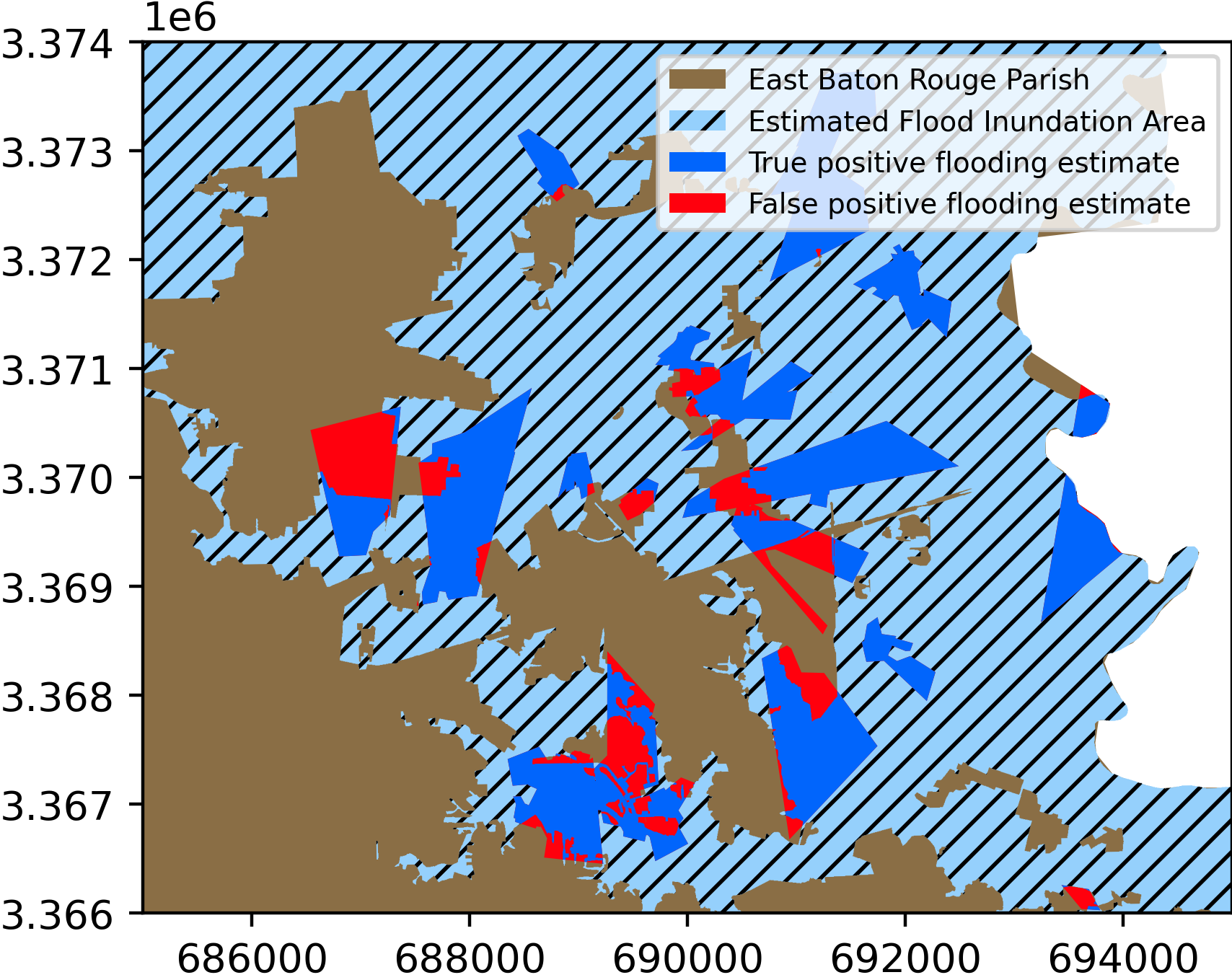}
         \caption{Our approach}
         \label{fig:damage_maps_closeup}
     \end{subfigure}
     
     \caption{Close-up of flooding estimates for image footprint and our approach, showing true and false positive regions.}
     \label{fig:closeup}
\end{figure}

Figure~\ref{fig:closeup} provides a closeup comparison of the footprint (SfM+binary classification) and (SfM+CAM) approaches. Precision is the total area colored dark blue divided by the sum of the dark blue and red areas.
\label{app:closeup}

\end{document}